\documentclass[fleqn,usenatbib]{mnras}
\usepackage{newtxtext,newtxmath}
\newcommand{\peijin}[1]{{#1}}
\usepackage{hyperref}

\usepackage{tikz}
\usetikzlibrary{positioning, shapes.geometric,matrix}
\usepackage[T1]{fontenc}

\DeclareRobustCommand{\VAN}[3]{#2}
\let\VANthebibliography\thebibliography
\def\thebibliography{\DeclareRobustCommand{\VAN}[3]{##3}\VANthebibliography}

\usepackage{graphicx}	
\usepackage{amsmath}	

\title[ConvRFI]{RFI Flagging in Solar and Space Weather Low Frequency Radio Observations}

\author[Zhang et al.]{Peijin Zhang$^{1,2,3}$\thanks{E-mail: peijin@nao-rozhen.org},
Andr\'e R. Offringa $^{2}$,
Pietro Zucca $^{2}$,
Kamen Kozarev $^{1}$,
Mattia Mancini $^{2}$,
\\
$^{1}$ Institute of Astronomy and National Astronomical Observatory, Bulgarian Academy of Sciences, Sofia 1784, Bulgaria\\
$^{2}$ ASTRON, The Netherlands Institute for Radio Astronomy, Oude Hoogeveensedijk 4, 7991 PD Dwingeloo, The Netherlands\\
$^{3}$ Astronomy \& Astrophysics Section, Dublin Institute for Advanced Studies, Dublin 2, Ireland\\
}

\date{Accepted XXX. Received YYY; in original form ZZZ}

\pubyear{2022}

\begin{document}
\label{firstpage}
\pagerange{\pageref{firstpage}--\pageref{lastpage}}
\maketitle

\begin{abstract}
Radio spectroscopy provides a unique inspection perspective for solar and space weather research, which can reveal the plasma and energetic electron information in the solar corona and inner heliosphere. However, Radio-Frequency Interference (RFI) from human activities affects sensitive radio telescopes, \peijin{and} significantly affects the quality of observation. Thus, RFI detection and mitigation for the observations is necessary to obtain high quality, science-ready data. The flagging of RFI is particularly challenging for the solar and space weather observations at low frequency, because the solar radio bursts can be brighter than the RFI, and may show similar temporal behavior. In this work, we investigate RFI flagging methods for solar and space weather observations, including a strategy for AOFlagger, and a novel method that makes use of a morphology convolution. These algorithms can effectively flag RFI while preserving solar radio bursts. 
\end{abstract}



\section{Introduction}

High-resolution solar and space weather radio spectroscopy can provide rich information for the study of transient short-term radio bursts, the density fluctuations of the  inner heliosphere and ionosphere detection, as well as long-term variations. Solar activity and its interaction with the plasma of the inner heliosphere generates radio emissions, which can be used to characterize the energy-releasing process and the plasma properties in the solar atmosphere, as well as in interplanetary space. For example, tied-array-beam observations were used to study in detail dynamic spectrum observations of Type III solar radio bursts \citep{zhang2019source}. The fine structures in a solar Type II radio burst can indicate the details of the shock evolution in a solar eruption \citep{magdalenic2020fine}. Also, the interplanetary scintillation (IPS) method uses the spectroscopy of a static astronomical source (e.g., Cygnus A, Cassiopeia A) to diagnose the small-scale density structures in the inner heliosphere and ionosphere from electron density fluctuation in the dynamic spectra. 

The Low-Frequency Array (LOFAR) \citep{van2013lofar} has proven to be a useful instrument for the study of solar activity and space weather \citep{oberoi2004lofar,dkabrowski2016prospects}. The project `LOFAR for Space Weather' \citep[LOFAR4SW]{carley2020radio} aimed at providing an upgraded design to make LOFAR capable of performing continuous observations of the Sun and space weather during the possible time window. Recently, a new project led by Dr. Pietro Zucca at ASTRON started, `Incremental Development Of LOFAR Space weather' (IDOLS), IDOLS \peijin{creates} all-time single station observation \peijin{of} the Sun and space weather and a daily interferometric image of the Sun with \peijin{the array of both Low Band Antenna (LBA), and High Band Antenna (HBA)}. For the dynamic spectrum observation, IDOLS uses a single core station (ID:CS032) detached from the LOFAR network to perform solar observations during the day and scintillation observations during the night. 
With the current development of IDOLS, a foreseeable increasing amount of spectroscopy data will be available for solar and space weather monitoring and research work.

Unfortunately, radio frequency interference (RFI) can significantly damage the quality of observations, especially in the low frequency range (<300MHz) \citep{offringa2013lofar}.
RFI signal contamination is a long-standing problem in radio astronomy, and is becoming more challenging because of the increased radio spectrum occupancy by technology. RFI flagging has been widely studied and many different algorithms have been developed. 
\texttt{AOFlagger}  \citep{offringa2012morphological} is a flagger based on filtering, combinatorial thresholding and morphological operations. \texttt{AOFlagger} is highly efficient and widely used in low-frequency radio observations, and is used in the default pipeline of LOFAR interferometric data preprocessing of imaging observations. However, it is less commonly used for transient processing. 
Recently, machine learning techniques have been applied for RFI detection, \citep{zhang2019source}. \cite{yang2020deep} proposed a machine learning model called RFI-Net. The method is implemented into the RFI flagging of the Five-hundred-meter Aperture Spherical radio Telescope (FAST). \cite{sun2022robust} developed an RFI-flagging tool based on Convolutional Neural Networks (CNN). The machine learning methods usually requires large training \peijin{ datasets}, and the flagging running process is usually computationally intensive.

The RFI flagging task for solar and space weather observations is different from the flagging for imaging observations with long integrations (e.g., deep extragalactic surveys). For such imaging observation, the signal of the target object for imaging is usually static or slow varying. Therefore, threshold-based methods are effective in marking the very bright and fast varying RFI. For example, the \texttt{SumThreshold} step in \texttt{AOFlagger} \citep{offringa2010post},  could effectively identify the samples with a high probability  to be RFI.
Then, the scale-invariant rank (SIR) operation is used to detect temporally and spectrally nearby weaker RFI \citep{offringa2012morphological,van2016efficient}.
These signal characteristics differs from solar and space weather observations, where the solar radio burst can be strong (sometimes much brighter than the RFI), and changes rapidly in both time and frequency. 
A flagging strategy designed for non-transient observations will classify the samples with solar radio bursts as RFI.

This work addresses the problem of RFI flagging in the solar and space weather observations. We propose a flagging method based on the morphological feature matching in the dynamic spectrum, an updated flagging strategy of \texttt{AOFlagger}, and a hybrid method combining the first two methods. We also test the flagging precision for these methods. 

This paper is arranged as follows: in Section \ref{S2}, we present the algorithms and how they are implemented in the data processing of LOFAR solar and space weather observations. In Section \ref{S3}, we show the result of the flagging in the observation data and simulation dynamic spectra. Section \ref{S4} presents the implementation of the method. A summary and discussion are given in Section \ref{S5}.

\section{Algorithm}
\label{S2}

The dynamic spectrum data processing for solar and space weather observations of LOFAR includes two steps: flagging and averaging. 
The purpose of flagging and averaging for the solar and spaceweather spectroscopy is to obtain a smoother and clearer dynamic spectrum to study the spectrum features (e.g., the frequency drift rate of solar Type III radio bursts, or the bandwidth and duration of solar noise storms). 
The flagging step will create a binary map that marks the RFI samples, and the averaging step will downsample the dynamic spectrum to a lower resolution and smaller size, to ease the data distribution and transferring. 
The averaging will discard the samples marked as RFI. This section presents the details of the data processing steps for the dynamic spectrum of solar and space weather observations.

\subsection{Flagging}
In this sub-section, we introduce the RFI identification method.
\subsubsection{ConvRFI}
The idea of using a morphological convolution for RFI detection, is based on the description of the RFI morphology (or shape) with convolution cores, and applying the convolution cores to the convolution operation for the dynamic spectrum. In the convolution result array, the pixels which match the corresponding morphology described by the kernel will be positive, representing RFI detected.

We select convolutional kernels that are designed to match the typical behavior of RFI. In observations, the most common types of RFIs are local lightning storms and communication transmissions. The spectrum of lightning is wide-band and transient. It appears as \peijin{features parallel to the frequency} axis in the dynamic spectra; while the spectrum of communication transmissions is narrow-band, appearing \peijin{parallel to the time} axis in the dynamic spectra. Considering these characteristics of RFIs, the kernel is prepared as shown in Fig \ref{fig:kernel}. Each kernel ($K_i$) is a binary value 2D array. Each element is set to either the value -1, or to $a_i$, which is a parameter set to a value larger than zero that represents the sensitivity of the kernel for flagging. Larger values of $a_i$ correspond to a higher sensitivity of flagging.

The flagging scheme $B_{\rm flag\it}$ can be expressed as:
\begin{equation}
    B_{\rm flag\it}=  B_{f0} \cup B_{f1} \cup B_{f2} \cup B_{f3} \cup B_{f4} \cup B_{f5}
    \label{eq:main}
\end{equation}
in which,
\begin{equation}
    \begin{aligned}
    B_i& = h(conv(D,K_i)) \quad [i=0,1,2,3,4,5]\\
    B_{f0}& = corr(B_0,h(K_0))\\
    B_{f1}& = corr(B_1,h(K_1))\\
    B_{f2}& = corr((B_2 \cap \overline{B_0}),h(K_2))\\
    B_{f3}& = corr((B_3 \cap \overline{B_0}),h(K_3))\\
    B_{f4}& = corr((B_4 \cap \overline{B_1}),h(K_4))\\
    B_{f5}& = corr((B_5 \cap \overline{B_1}),h(K_5))
    \end{aligned}
    \label{eq:conv}
\end{equation}
where $D$ is the dynamic spectrum, $K_i$ are the kernels shown in Fig. \ref{fig:kernel}, $B_i$ are the  convolution results with kernel $K_i$, indicating the  detection points, $B_{fi}$ are the binary-value arrays of flagging corresponding to the morphology described be kernel $K_i$, $\overline{B_{i}}$ is the logical negation of ${B_{i}}$, and $h(x)$ is the Heaviside function. $conv$ and $corr$ are the convolution and correlation operations defined as:
\begin{equation}
    \label{eq:conv_def}
    \begin{aligned}
        conv&(D,K)[i,j] = \\
     &\sum^{m=N}_{m=0} \sum^{n=N}_{n=0}  D\left[i-\left(m-\frac{N-1}{2}\right),j-\left(n-\frac{N-1}{2}\right)\right] K[m,n]
    \end{aligned}
\end{equation}
\begin{equation}
    \begin{aligned}
        corr&(D,K)[i,j] = \\
     &\sum^{m=N}_{m=0} \sum^{n=N}_{n=0}  D\left[i+\left(m-\frac{N-1}{2}\right),j+\left(n-\frac{N-1}{2}\right)\right] K[m,n]
    \end{aligned}
\end{equation}
with $N$ as the dimension of the kernel, $K_{0-1}$ is $N=3$, and  
$K_{2-5}$ is $N=5$.

\begin{figure}
    \centering
    \tikzset{
  matrixstyle/.style={
    matrix of math nodes,
    nodes={minimum size=0.5cm, {\ifdim\d=0pt \else draw\fi}, outer sep=0pt,inner sep=0,line width=\d},
}}
\def\d{0pt}
\ifdim\d=0pt 
    \def\DrawFlag{}%
\else
    \def\DrawFlag{draw}%
\fi

\begin{tikzpicture}
        \begin{scope} [scale = 0.92, every node/.style = {scale = 0.92}]
            \matrix (M11) [matrixstyle,left delimiter=(,right delimiter=)]{
                -1 & a_0 & -1  \\
                -1 & a_0 & -1 \\
                -1 & a_0 & -1 \\
            };
            \node[above=0pt of M11]{ $K_0$  };
            \draw[rounded corners,ultra thick, draw=black, fill=orange, opacity=0.2] (M11-3-2.south west) rectangle (M11-1-2.north east);  
            
            \matrix (M12) [matrixstyle,left delimiter=(,right delimiter=), right = 35pt of M11]{
                -1 & -1 & -1  \\
                a_1 & a_1 & a_1 \\
                -1 & -1 & -1 \\
            };
            \node[above=0pt of M12]{ $K_1$  };
            \draw[rounded corners,ultra thick, draw=black, fill=orange, opacity=0.2] (M12-2-1.south west) rectangle (M12-2-3.north east);  
            
             \matrix (M13) [matrixstyle,left delimiter=(,right delimiter=), right = 25pt of M12]{
                a_2 & a_2 & a_2 & -1 &-1\\
                a_2 & a_2 & a_2 & -1 &-1\\
                a_2 & a_2 & a_2 & -1 &-1\\
                a_2 & a_2 & a_2 & -1 &-1\\
                a_2 & a_2 & a_2 & -1 &-1\\
            };
            \node[above=0pt of M13]{ $K_2$  };
            \draw[rounded corners,ultra thick, draw=black, fill=orange, opacity=0.2] (M13-5-1.south west) rectangle (M13-1-3.north east);  
            
            \matrix (M23) [matrixstyle,left delimiter=(,right delimiter=), below = 20pt of M13]{
                a_3 & a_3 & a_3 & a_3 & a_3 \\
                a_3 & a_3 & a_3 & a_3 & a_3 \\
                a_3 & a_3 & a_3 & a_3 & a_3 \\
                -1 & -1 & -1 & -1 &-1\\
                -1 & -1 & -1 & -1 &-1\\
            };
            \node[above=0pt of M23]{ $K_5$  };
            \draw[rounded corners,ultra thick, draw=black, fill=orange, opacity=0.2] (M23-3-1.south west) rectangle (M23-1-5.north east);  
            
            \matrix (M22) [matrixstyle,left delimiter=(,right delimiter=), left = 12pt of M23]{
                -1 & -1 & -1 & -1 &-1\\
                -1 & -1 & -1 & -1 &-1\\
                a_3 & a_3 & a_3 & a_3 & a_3 \\
                a_3 & a_3 & a_3 & a_3 & a_3 \\
                a_3 & a_3 & a_3 & a_3 & a_3 \\
                };
            \node[above=0pt of M22]{ $K_4$  };
            \draw[rounded corners,ultra thick, draw=black, fill=orange, opacity=0.2] (M22-5-1.south west) rectangle (M22-3-5.north east);  
            
            \matrix (M21) [matrixstyle,left delimiter=(,right delimiter=), left = 12pt of M22]{
                -1 & -1 & a_2 & a_2 & a_2  \\
                -1 & -1 & a_2 & a_2 & a_2 \\
                -1 & -1 & a_2 & a_2 & a_2  \\
                -1 & -1 & a_2 & a_2 & a_2  \\
                -1 & -1 & a_2 & a_2 & a_2  \\
                };
        \end{scope}
            \node[above=0pt of M21]{ $K_3$  };
            \draw[rounded corners,ultra thick, draw=black, fill=orange, opacity=0.2] (M21-5-3.south west) rectangle (M21-1-5.north east);  

\end{tikzpicture}
    \caption{Convolution kernels. $a_{0-3}$ control the sensitivity of RFI detection.}
    \label{fig:kernel}
\end{figure}

Eq. (\ref{eq:conv}) shows that, if kernels $K_{2-5}$ convoluted with a very bright line, they will result \peijin{in} RFI-positive samples, meaning the strong line-like features marked by $K_{0,1}$ will  also be flagged as an edge-like feature by $K_{2-5}$.
When the lines are flagged as sharp edges, the samples near the bright lines will also be marked as RFI.
This will cause over-flagging near the strong line features if we directly take the union set of the flagging result of the line features and the sharp edges.
Thus, we have adopted the strategy to first perform the line feature convolution to flag the line features ($B_{f0,1}$), and then run the sharp edge features convolution excluding the line feature points, as expressed in Eq. (\ref{eq:conv}), $B_{f2}-B_{f5}$.

With a dynamic spectrum $D$ as input, after the operation described in Eq. (\ref{eq:main}), we can have a binary map of flagging $B_{\rm flag\it}$, with RFI pixels set to 1, and the rest of the points set to 0. The sensitivity parameters $a_i$ can be used to control the sensitivity of each morphology; for example, a large value of $a_0$ and a small value of $a_1$ makes the algorithm more sensitive to features \peijin{parallel to the time} axis  but less sensitive to features \peijin{parallel to the frequency} axis.

The most computationally intensive tasks in the  flagging steps are the convolution and correlation operations on 2D arrays. For this, we use the \texttt{Conv2D} \peijin{implemented} in \texttt{PyTorch}, \peijin{which is a machine learning framework providing a highly optimized implementation of convolution operations available on both CPU and GPU}. The performance and resource requirements are presented in detail in Sec \ref{sec:performance}.

\subsubsection{AOFlagger Local-RMS}

The RFI selection of \texttt{AOFlagger} is based on weighted low-pass filtering; combinatorial thresholding (\texttt{SumThreshold}) and morphological expansion (SIR operator).
The thresholding step is sensitive to large values; it is thus likely to select all samples during bright solar radio bursts. To achieve RFI flagging for dynamic spectra with strong radio bursts, we use relative thresholding in reference to the root-mean-square (RMS) of nearby samples, the RMS is weighted by a Gaussian kernel excluding the flagged samples.
The local-RMS method can help avoid flagging bright samples in the radio bursts as RFI.

A further option is to use the detections results of \texttt{ConvRFI} as the initial flags for the \texttt{local-RMS AOFlagger} strategy. This will be referred to as the `hybrid method' in the following sections.

\subsection{Averaging}

After flagging, to down-sample the data into a smaller size, one can perform the averaging with the following expression:
\begin{equation}
    \label{eq:avg}
    I_{avg}[n,m] =  \frac{\sum^{N-1}_{i=0} \sum^{M-1}_{j=0} (I_{raw}\times\overline{B_{flag}})[n\times N+i,m\times M+j]}{\sum^{N-1}_{i=0} \sum^{M-1}_{j=0} \overline{B_{flag}}[n\times N+i,m\times M+j]}
\end{equation}
where the multiplication and division operation in this equation are element-wise, and $I_{avg}[n,m]$ is the averaged dynamic spectrum at time index $n$ and frequency index $m$. Within $\overline{B_{flag}}$, the RFI positive samples corresponds to value `0', and the RFI negative samples corresponds to value `1' in the result array. Thus, with this method, the flagged samples are treated as 0-weight in the averaging.
The averaging window size is $N$ and $M$ in time and frequency, respectively. As shown in Eq. \ref{eq:avg}, the averaging step and averaging window are of the same length.

The averaging described in Eq. (\ref{eq:avg}) can be expressed in the form of a convolution operation:
\begin{equation}
    I_{avg} =  \frac{conv(I_{raw}\times\overline{B_{flag}},U_{MN}; \rm stride=\it [M,N])}{conv(\overline{B_{flag}},U_{MN}; \rm stride=\it [M,N])},
    \label{eq:conv_avg}
\end{equation}
Where $U_{MN}$ is a matrix with all elements set to 1 and a size of $M\times N$,  stride-length defines how many steps we take when sliding the convolution core (e.g. $U_{MN}$) across the array (e.g. $\overline{B_{flag}}$). This convolution operation can also be implemented with the \texttt{Conv2D} module in \texttt{PyTorch}.

With the above flagging and averaging methods, the procedure is : flagging in high resolution and then averaging down to low resolution.
By flagging in the higher resolution input data, the RFI (narrow in time and frequency) would contaminate a smaller portion of the dynamic spectrum.
As a result, the averaging result will have more contribution from the RFI uncontaminated sample.


\section{Flagging Results}
\label{S3}

In this section, we present the test results of the various methods described in the previous section.

\subsection{Dataset}

The algorithms presented above are designed and tested for high resolution dynamic spectra at low frequency (10-90~MHz). We use solar observations with the low-band antennas (LBA) of LOFAR. The time resolution of the raw data is 10.5~ms. The observation covers 10-90 MHz with 6400 frequency channels, giving a frequency resolution of 12.2 kHz. Values are stored as single precision floating point values. The data rate is 8.19~GB/h, saved in HDF5 files. 
For all-day spectroscopy observations as part of the project IDOLS, these observations will generate approximately 0.7~TB of full Stokes (I,Q,U,V) data.

\begin{figure*}
    \centering
    \includegraphics[width=0.77\linewidth]{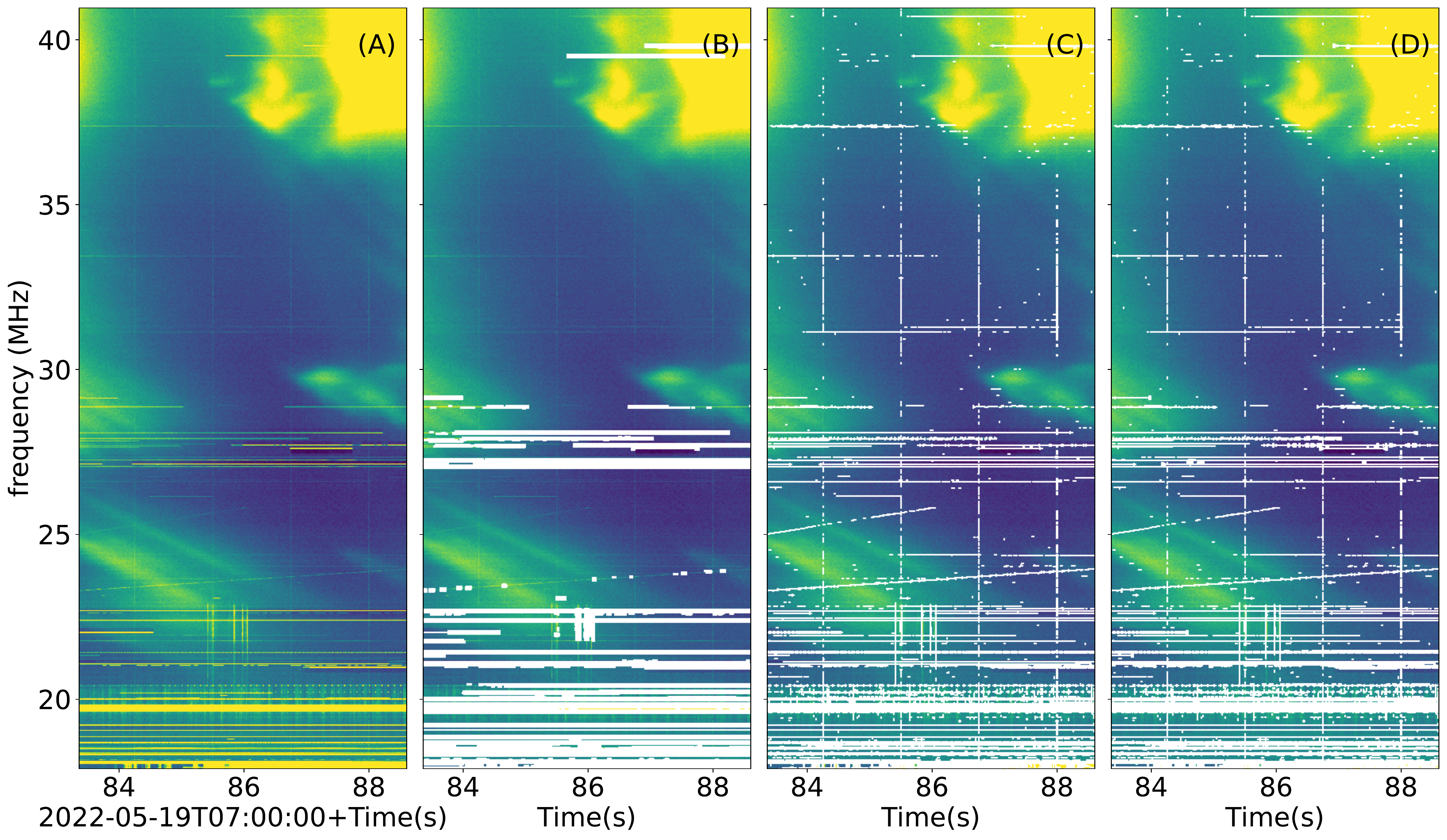}
    \caption{\texttt{ConvRFI} applied to the observed dynamic spectrum with different parameter combinations. The white patches indicate the mask of flagged time-frequency samples.
    Panel (A) presents the dynamic spectrum without flagging, 
    panel (B) shows the result of sharp edge detection; panel (C) shows the horizontal and vertical line detection; and panel (D) uses a combination.}
    \label{fig:result}
\end{figure*}

To test our method, we use a dynamic spectrum recorded with \peijin{the} LBA on 19-May-2022 near 07:00~UT. The raw dynamic spectrum is shown in Fig. \ref{fig:result}(A). The dynamic spectrum shows a strong solar radio burst, as well as strong RFI.

\begin{figure*}
    \centering
    \includegraphics[width=0.88 \linewidth]{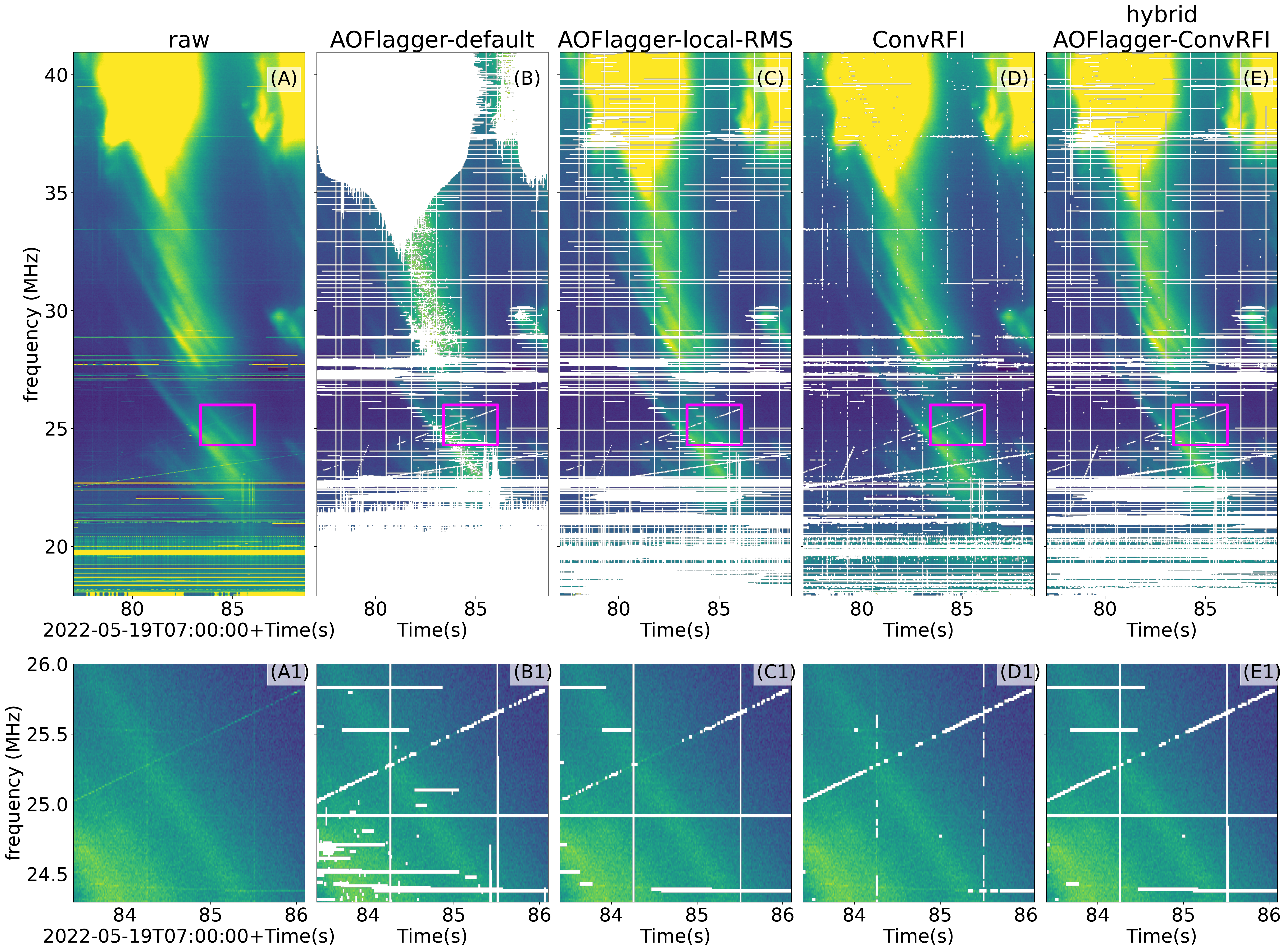}
    \caption{Result of various flagging strategies. Panel (A) is the observed dynamic spectrum, panels (B-E) are the flagging results, Panels (A1-E1) zoom in on the magenta rectangles in panels (A-E).}
    \label{fig:AOFlagger}
\end{figure*}

\subsection{Flagging on observed RFI}

Fig. \ref{fig:result} (B,C,D) shows the flagging results (masks) of \texttt{ConvRFI} with different parameter combinations. Panel (B) uses $a_i=$~[0.01,0.01,0.45,0.45],  which flags mainly on sharp edge features without triggering specifically on line-like features. Panel (C) uses $a_i=$~[1.66,1.66,0.01,0.01],  which flags mainly on line-like features without triggering specifically on sharp-edge features. Panel (D) uses $a_i=$~[1.66,1.66,0.45,0.45], which combines sharp edge detection and line detection. Comparing the three parameter combinations, we can see that the flagging result is best when $a_i=$~[1.66,1.66,0.45,0.45] is used (panel D).
If we compare panel (B) and panel (D) of Fig. \ref{fig:result}, for frequencies near 29~MHz and 40~MHz, we can clearly see the over-flagging of the sharp edge feature without line feature flagging.
The result of using combined parameters (panel D) shows accurate flagging of line and sharp edge features. In our pipeline, we therefore use the parameter combination $a_i=$[1.66,1.66,0.45,0.45] for LOFAR-LBA flagging.

As a comparison, we also applied \texttt{AOFlagger} using the default flagging strategy without further parameter tweaking, as well as the \texttt{AOFlagger local-RMS} method. As shown in Fig. \ref{fig:AOFlagger}, we can see that, with the default flagging strategy of \texttt{AOFlagger} in panel (B), the RFI pixels are well-flagged, but a significant part of the solar radio bursts is also flagged out as RFI. 
This demonstrates that the RFI detection of high-time-resolution data, which may contain transients of interest, requires the design of different methods.

One method that we test, is to adapt the \texttt{aoflagger} strategy so that the flag thresholds of the \texttt{SumThreshold} step are relative to the local RMS. We will refer to this method as \texttt{AOFlagger local-RMS}. With this strategy, most samples of the solar radio burst are preserved (not flagged), as shown in panel (C) of Fig. \ref{fig:AOFlagger}.
Comparing the results of \texttt{AOFlagger local-RMS} and \texttt{ConvRFI}, we can see that the vertical RFI lines are better detected in \texttt{AOFlagger local-RMS}, but not fully masked in \texttt{ConvRFI}. \texttt{ConvRFI} tends to ignore weak RFI within the solar radio bursts, but preserves more radio burst samples.
As shown in Panel (C1,D1,E1), 
by using a hybrid method where the flag output of \texttt{ConvRFI} is used as input for the \texttt{AOFlagger local-RMS} method, the result has a better flagging coverage on both the slash-line-shaped RFI, and the narrowband RFI.


\begin{figure}
    \centering
    \includegraphics[width=0.98\linewidth]{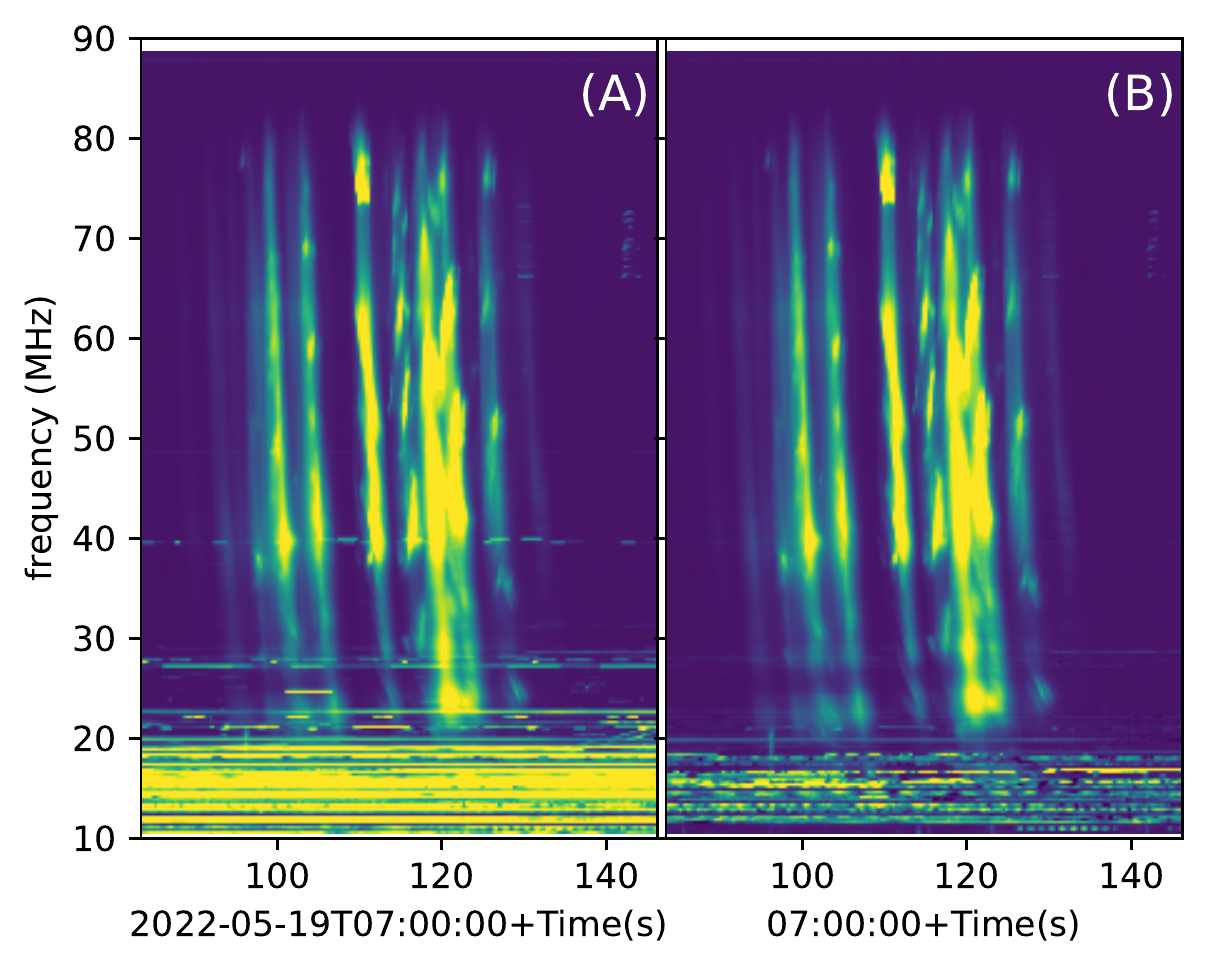}
    \caption{The averaged dynamic spectrum. Panel (A): without flagging; panel (B) with flagging.}
    \label{fig:avg}
\end{figure}

\begin{figure}
    \centering
    \includegraphics[width=0.95\linewidth]{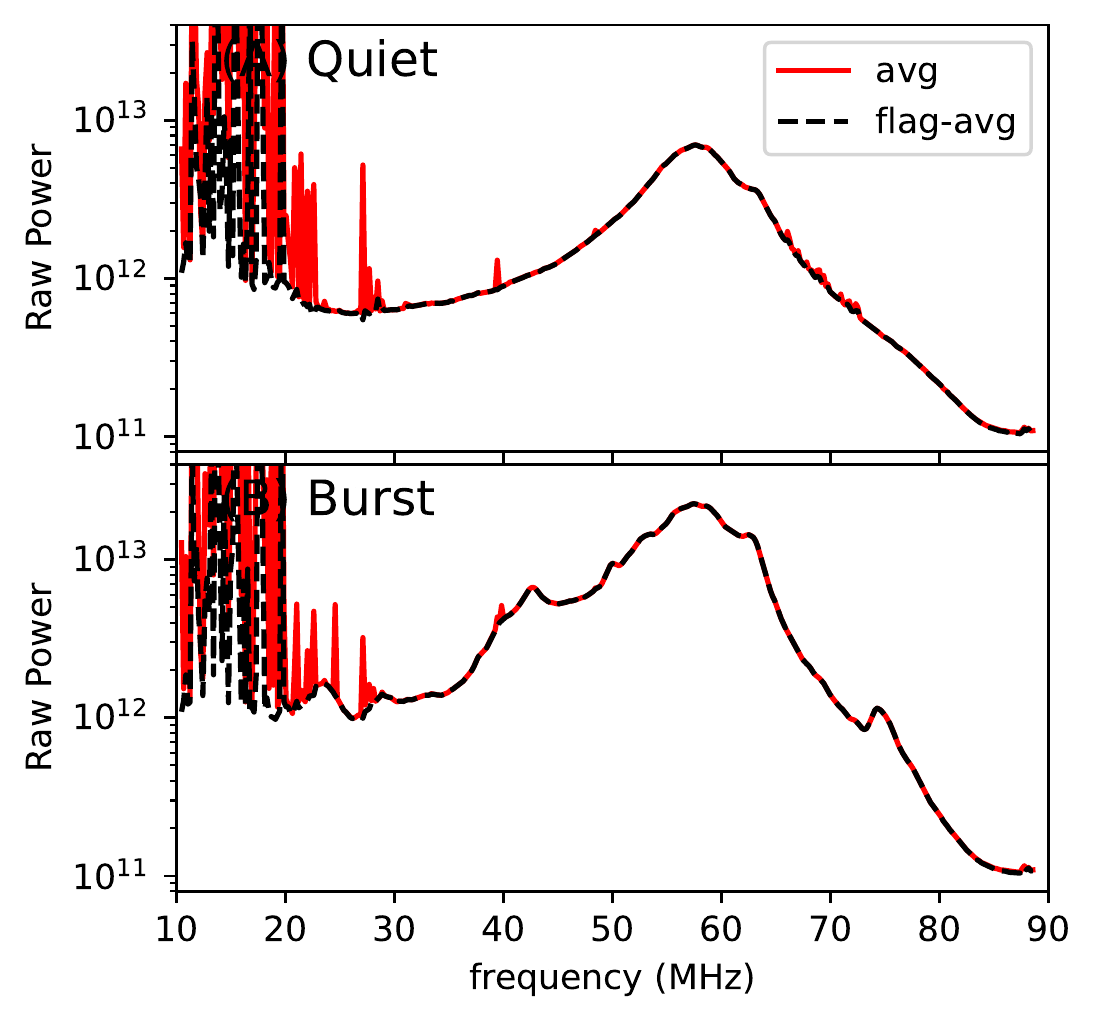}
    \caption{The raw power spectra of the quiet time and also the burst time.}
    \label{fig:bandpass}
\end{figure}

Fig. \ref{fig:avg} shows the averaged dynamic spectrum with and without flagging. 
The averaging window of time and frequency is 64 time samples and 16 frequency channels, respectively. The RFI between 20-30~MHz and at 40~MHz is reduced, and the effect of RFI on the radio burst is reduced as well.  
Fig. \ref{fig:bandpass} shows power spectrum of raw data taken from 15~s integral from an interval around the burst and from a quiet interval. Before flagging, the spectrum shows residual spikes due to RFI in both intervals. After flagging,  the power \peijin{spectrum} is smoother. 

\subsection{Evaluation}

\begin{figure}
    \centering
    \includegraphics[width=0.85\linewidth]{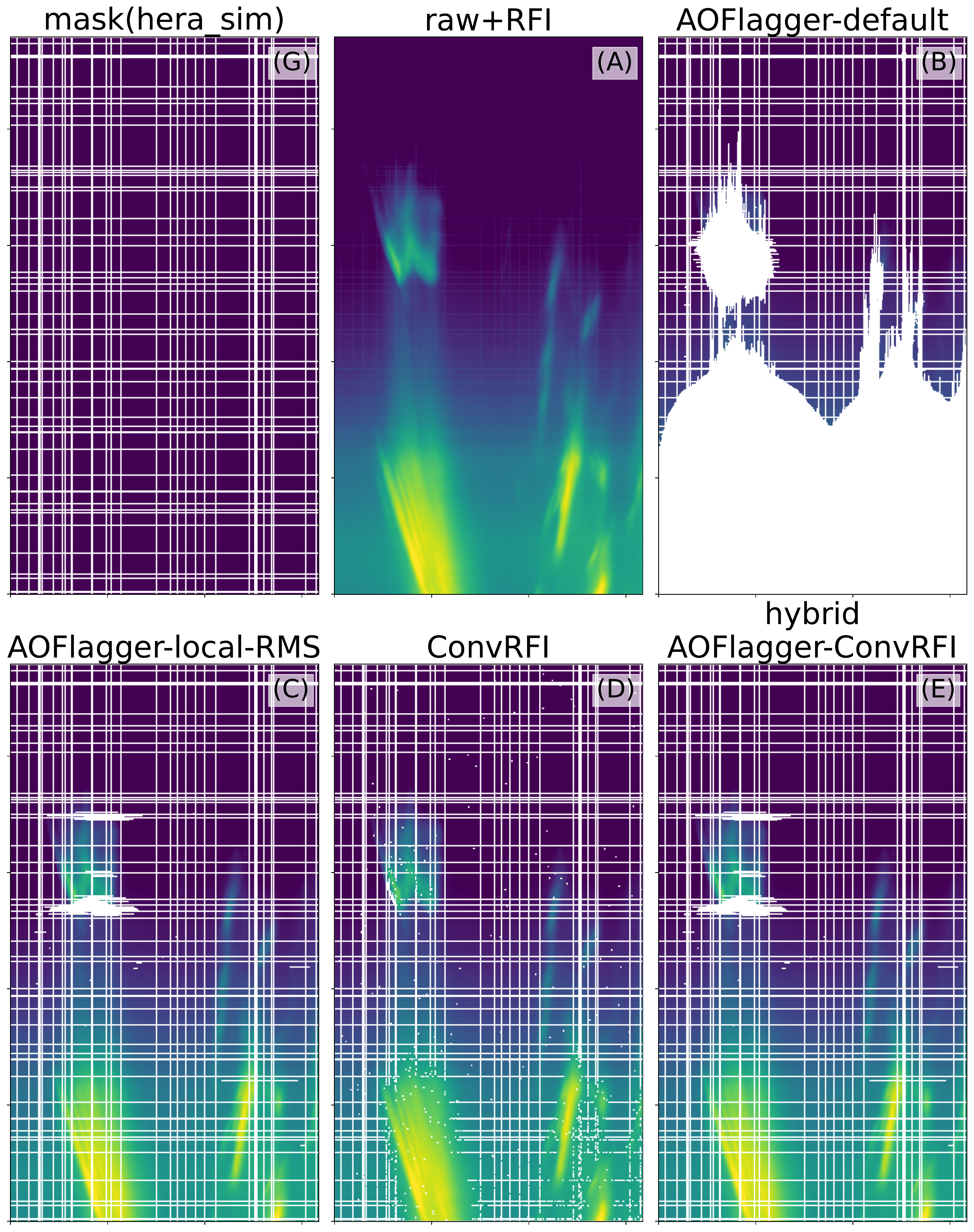}
    \caption{The RFI simulation overlapped on the observed radio bursts. Panel (G) shows the ground truth mask of flagging. Panel (A) shows the overlapped dynamic spectrum, panels (B, C, D, E) are the flagging results of four flagging methods indicated in the label.}
    \label{fig:hera_demo}
\end{figure}

To test the flagging performance of the methods using a ground truth, we overlap simulated RFI (\texttt{hera\_sim}) \citep{parsons2012per} to a segment of the  dynamic spectrum with no visible RFI. Fig. \ref{fig:hera_demo} shows an example of simulated RFI with intensity of 1\% of the peak intensity in the dynamic spectrum, and the flagging results of the four methods.
With the simulated RFI, we can compare the ground-truth RFI mask with the flagging result, from which the correctness of RFI flagging can be classified into four \peijin{categories}: \peijin{true positive (TP)} is the number of successfully detected RFI samples, \peijin{false positive (FP)} is the number of RFI-free samples flagged as RFI, \peijin{true negative (TN)} is the number of RFI-free samples not flagged as RFI,  \peijin{false negative (FN)} is the number of RFI samples not flagged.
The ratio of these parameters: 
\begin{equation}
    \rm{Recall}=\frac{\rm{TP}}{\rm{TP}+\rm{FN}}
\end{equation}
\begin{equation}
    \rm{Precision}=\frac{\rm{TP}}{\rm{TP}+\rm{FP}}
\end{equation}
\begin{equation}
    \rm{Accuracy}=\frac{\rm{TP}}{\rm{TP}+\rm{FN}+\rm{FP}}
\end{equation}
is used to evaluate the flagging results.

\begin{figure*}
    \centering
    \includegraphics[width=0.85\linewidth]{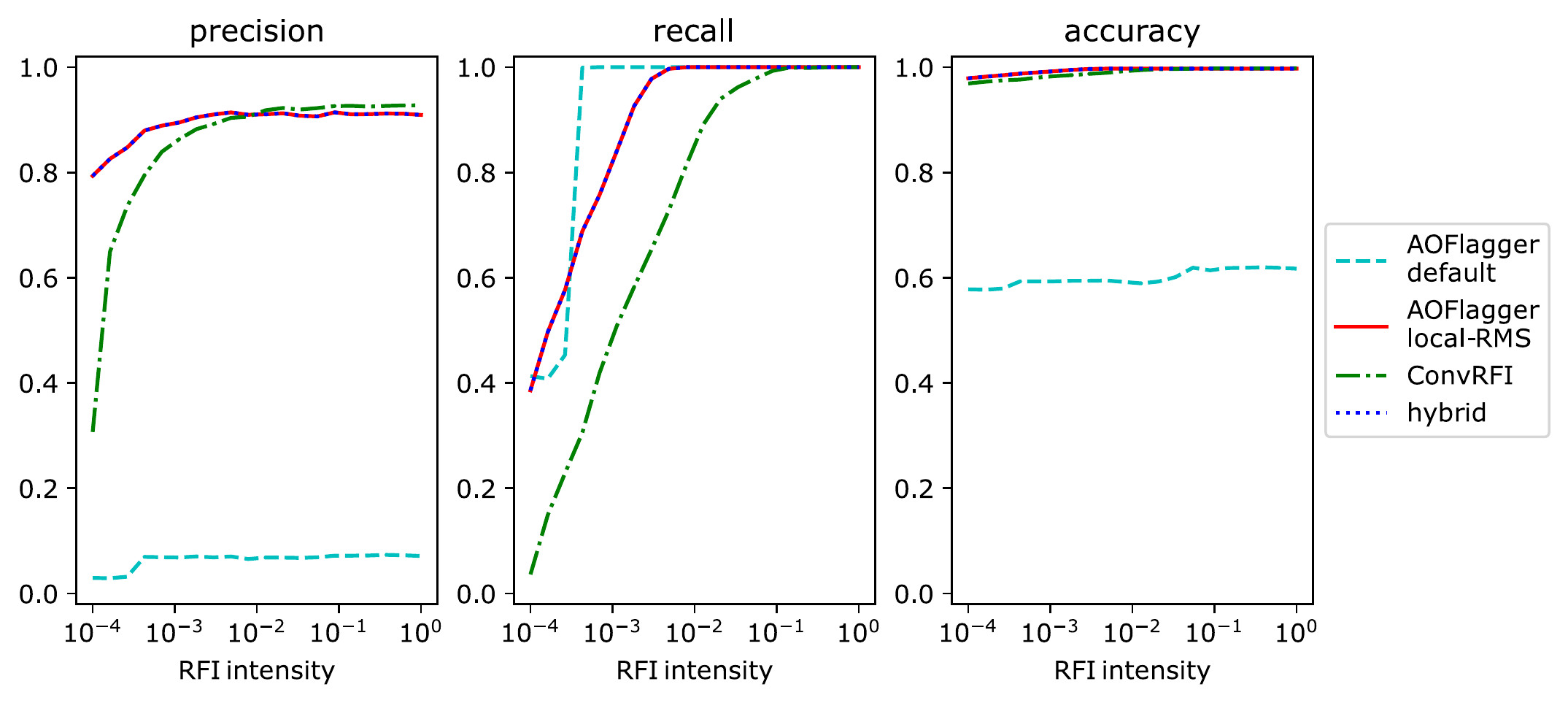}
    \caption{Evaluation results of the 4 methods.}
    \label{fig:evaluation}
\end{figure*}

We generate the RFI at different intensities to benchmark the performance of the flagging methods in different RFI conditions. \peijin{ The RFI relative intensity is defined as the intensity of the generated RFI normalized with the peak radio burst intensity. For example, RFI relative intensity=0.1 means the received RFI flux is 0.1 times the radio burst peak flux}. 
The resulting precision recall, and accuracy are shown in Fig. \ref{fig:evaluation}. 
We see that the \texttt{AOFlagger local-RMS} and \texttt{hybrid} methods have very similar results for all three measures. The \texttt{AOFlagger default} (not optimized) method has very low precision but high recall in the test result, representing a high over-flagging ratio. In terms of accuracy, the \texttt{AOFlagger local-RMS}, \texttt{ConvRFI} and \texttt{hybrid} methods all score above 0.96 for the  tested relative RFI intensity range ($10^{-4}\sim 1$).
As shown in the precision panel of Fig. \ref{fig:evaluation}, we can see that \texttt{AOFlagger local-RMS} and \texttt{hybrid} perform significantly better with weak RFI situation, while \texttt{ConvRFI} scores higher in stronger RFI (relative intensity > 0.01).

\subsection{Efficiency}
\label{sec:performance}

The computational efficiency of \texttt{AOFlagger}-based method is well described in \cite{offringa2012morphological}, and more recently in \cite{andre2023apertif}. \texttt{AOFlagger} is shown there to reach a flagging speed of 370 MB/s on an 8 core desktop machine. In this section, we provide a performance test of the \texttt{ConvRFI} method.

The flagging task mainly consumes three types of resources: input-output (IO), memory and computation. For dynamic spectrum data stored as HDF files, the IO time consumption refers to the time used to transfer data from disk to host memory (RAM), and the performance is determined by the disk reading speed. The memory size usage is the total size of the temporary variables used for flagging, (also for GPU computation). For GPU computations, an extra step of transferring data from host memory to GPU memory is added, which could also be time-consuming. The time consumption of the computation part is determined by the algorithm efficiency and the computing resource (cores of CPU or GPU).
To test the efficiency, we use a data segment of $0.32\times10^6$ time slots, with 6400 frequency channels, the total size is about 7.6~Gigabytes. We tested the algorithm on two types of machines: an 8-core CPU laptop with GPU and a server with 128 cores.

\begin{table*}
    \begin{tabular}{|c | llllll |} 
    \hline\hline
    [ms]($10^{-3}$s)& IO & RAM to VRAM & \texttt{ConvRFI} & VRAM to RAM & Averaging & Total    \\ \hline
    GPU RTX3060  & 5557.9 & 2549.5 & 5012.8 & 2937.8 & 5336.5 & 21394.5     \\
    CPU 8-core   & 5128.7 & - & 148420.3 & - & 4989.1 & 158538.1    \\
    CPU 128-core & 2332.4 & - & 32790.5 & - & 1111.1 & 36234.0 \\ 
    \hline
    \end{tabular}
    \caption{The time cost of \texttt{ConvRFI} to process the size of 7.6~GB data, with input data shape of 320000 time slots and 6400 frequency channels, 
    including the data reading (IO), the transfer time between CPU and GPU, and the step of averaging (downsampling) in both time and frequency.
    The output of execution is a binary map with the RFI time-frequency points as `1' and else as `0' and a small-size dynamic spectrum.}
    \label{tab:bench}
\end{table*}

The benchmark result is shown in Table \ref{tab:bench}. The first two rows represent runs on a laptop, and the third row is on a server. From the benchmark of the three rounds of processing, we can see that the GPU run has the best overall performance, although it requires an extra step of transferring data from host memory (RAM) to GPU memory (VRAM). For a data segment of 7.6~GB, the GPU version took 21.4~seconds in total, of which the time consumption for IO, flagging, and averaging took similar amounts of time - each about 5 seconds. The pure computational speed of flagging is about 1.5~GB/s.
For CPU versions with both 8 and  128 cores, the most time-consuming part is the flagging: the pure computational speed of flagging is about 0.05~GB/s on an 8-core CPU and 0.23~GB/s on an 128-core CPU. The IO performance is much better on the server due to higher-speed hard drives.

Overall, the flagging speed including the IO and averaging time can reach 0.35~GB/s on GPU-laptop, and 0.2~GB/s on CPU-server.

\subsection{RFI ratio}

\begin{figure}
    \centering
    \includegraphics[width=0.99\linewidth]{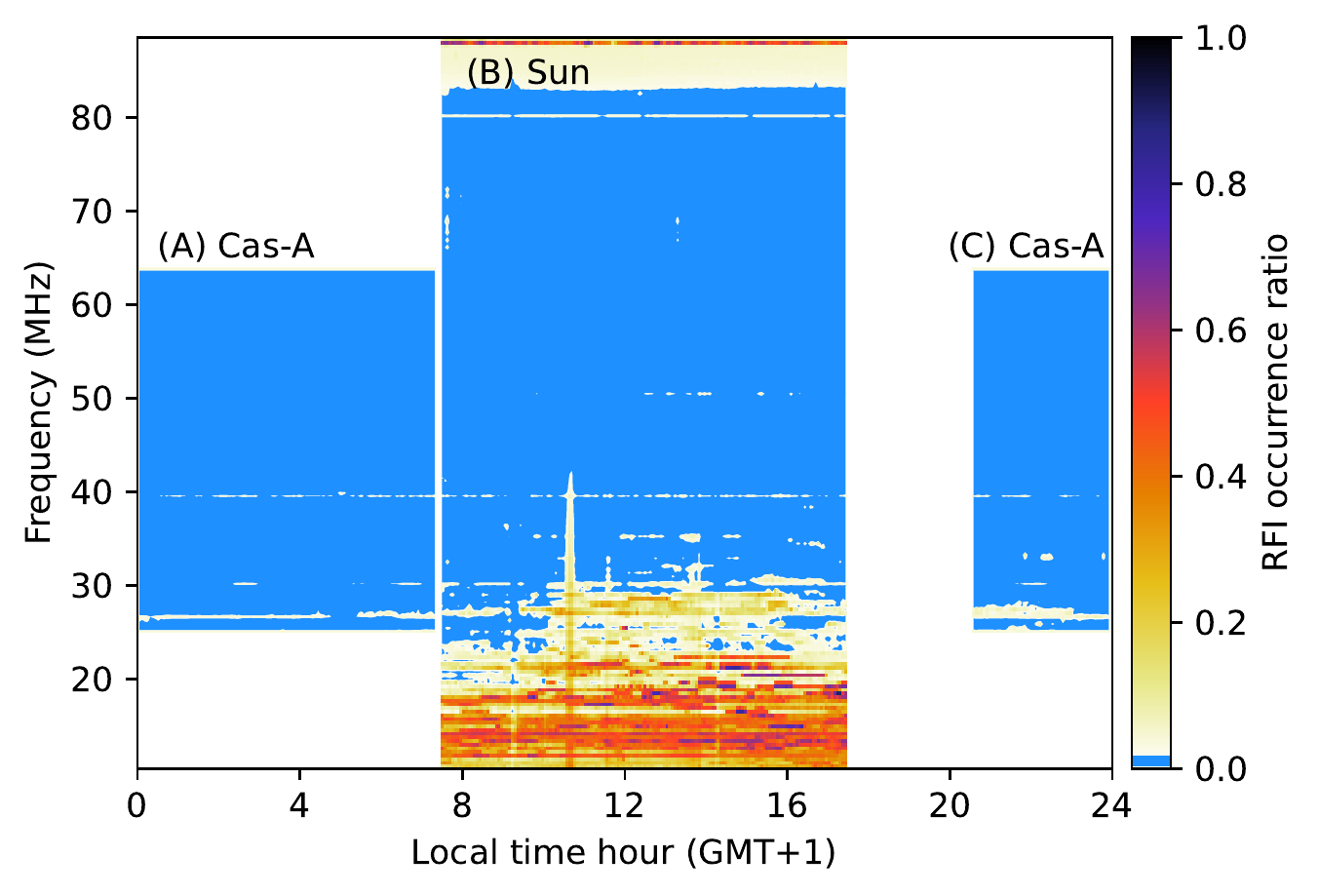}
    \caption{The RFI ratio of the averaged dynamic spectrum.}
    \label{fig:RFIratio}
\end{figure}

With the tested methods, most RFI can be detected and the spectrum of interest can be extracted, We can determine the residual spectral coverage after RFI detected results.
To assess the residual spectral coverage for the solar and space weather observations of LOFAR, we run flagging on the spectrum data of both daytime solar observations and nighttime ionospheric scintillation observations targeting Cassiopeia A, in this test we use \texttt{ConvRFI} for better precision with strong RFIs. 
The data was recorded by the LOFAR core station \textit{CS032LBA}. The nighttime observation is operating in the frequency range of 25-65~MHz, the daytime observation is operating in the frequency range of 10-90~MHz. As shown in Fig. \ref{fig:RFIratio}, in the majority part of the time-frequency domain, the RFI ratio is still below 2\% (indicated by the blue area), which is defined as `high quality dynamic spectrum'. In the lower frequency range (<30~MHz), the quality is much worse, reaching 10\%-20\% near 30MHz, and can reach >40\% below 20~MHz. there is persistent RFI near 28, 40, and 80~MHz. 

Comparing Fig. \ref{fig:RFIratio} and Fig. \ref{fig:avg}, for the frequency range of >30~MHz, the RFI ratio is mostly below 5\%, and the features of the radio burst can be well presented. In the frequency range of 20-30~MHz, the RFI ratio is about 5\%-30\%, and we can see some faint artifacts in the flagged and averaged dynamic spectrum. For the frequency range of <20~MHz, a major part of the spectrum has RFI ratio above 40\%, \peijin{due in part to ionospheric} reflection and absorption. \peijin{Therefore, most} of the solar radio bursts cannot be resolved below 20~MHz.

\section{Implementation}
\label{S4}

\begin{figure*}
    \centering
    \includegraphics[width=0.88\linewidth]{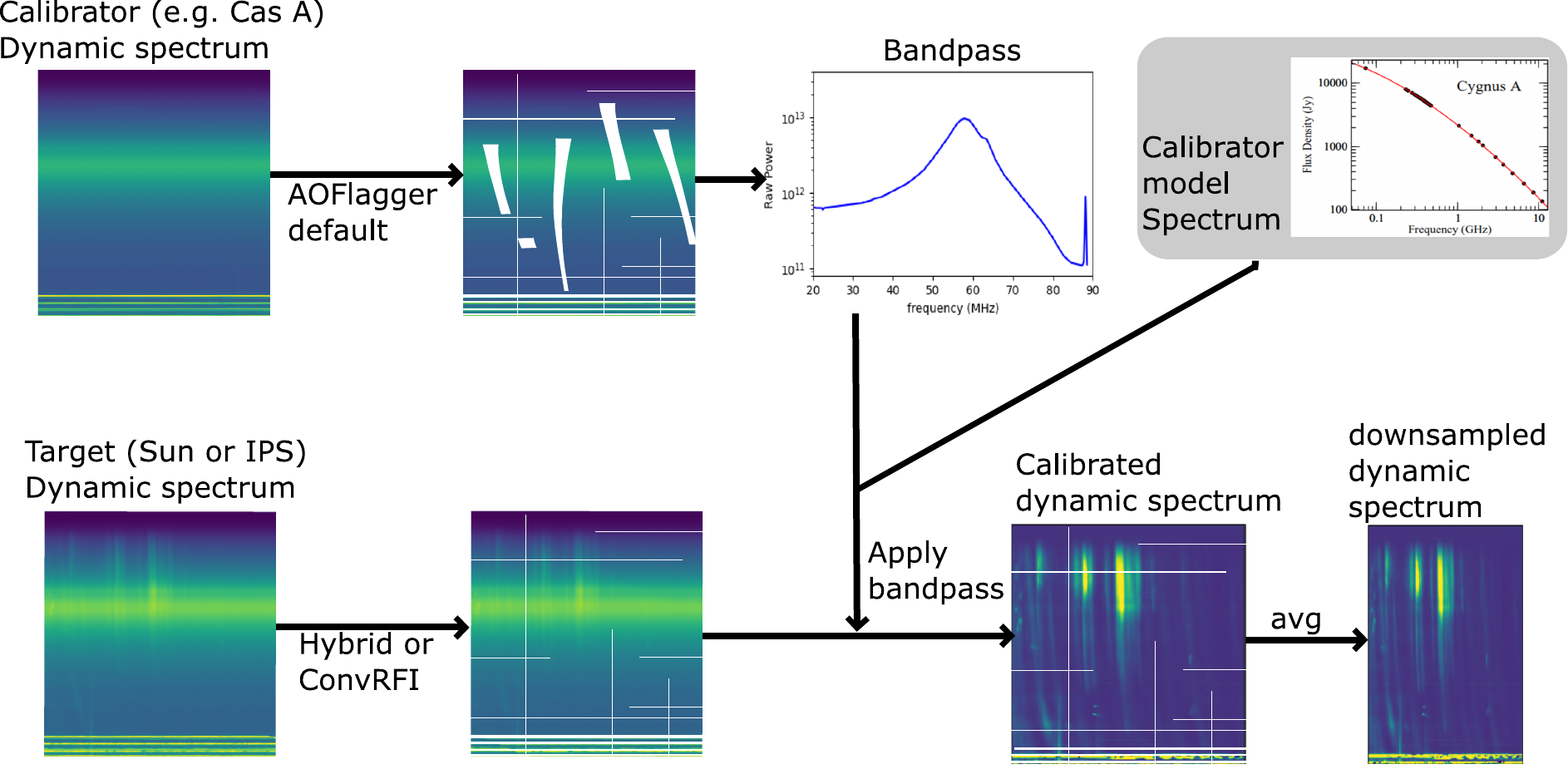}
    \caption{Concept flowchart of the data pre-processing pipeline for LOFAR solar and space weather observations.}
    \label{fig:pipeline}
\end{figure*}

We implement RFI detection in the pre-processing pipeline of the project `Incremental development of LOFAR for space weather' (IDOLS).
We select different methods for different scenarios.
The data processing procedure is shown in Fig. \ref{fig:pipeline}. As LOFAR is capable of simultaneous multi-beam observation \citep{mol2011lofar}, the telescope will provide simultaneous solar and calibrator observations. The calibrator observation is processed with the \texttt{AOFlagger default} pipeline, considering the calibrator is static in flux. It is then averaged in time to get the bandpass response of the calibrator ($\rm Obs_{cal}(\it f)$). Combining $\rm Obs_{cal}(\it f)$ with the spectrum flux model ($\rm Model_{cal}(\it f)$) of the calibrator (e.g. \citep{perley2017accurate}), we can apply the relative calibration for solar observation with
$$\rm Flux_{sun}(\it t,f) =\rm \frac{Model_{cal}(\it f)}{\rm Obs_{cal}(\it f)}\times \rm Obs_{sun}(\it t,f).$$
For the target observation, the flagging method depends on the objective.  \texttt{ConvRFI} is used for solar radio bursts, in which case weak RFI removal is not strongly required and \texttt{ConvRFI} has higher precision with strong RFI (shown in Fig. \ref{fig:evaluation}). The \texttt{hybrid} method is used for quiet Sun and fluctuations and the scintillation studies, which are not sensitive to dynamic spectrum completeness but sensitive to weak RFI.

\section{Summary and Discussion}
\label{S5}

In this work, we presented RFI flagging methods that are specifically aimed at low frequency solar and space weather observations with the LOFAR radio telescope. The three methods (\texttt{ConvRFI}, \texttt{AOFlagger local-RMS} and \texttt{hybrid})  perform well for dynamic spectra with solar radio bursts. The \texttt{ConvRFI} method is a simple and GPU enabled algorithm with a reasonable accuracy, that is therefore suitable as a preprocessing step. 
\texttt{AOFlagger local-RMS} and \texttt{hybrid} perform well for weak RFI situations, and are less likely to miss RFI samples.
Thus, \texttt{ConvRFI} is suitable for the pre-flagging and flagging tasks with more stringent requirements to retain the radio burst features, but lower requirements to flag all RFI samples. While the \texttt{AOFlagger local-RMS} and \texttt{hybrid} is suitable for the flagging task of RFI sensitive cases, with higher priority of flagging out RFI samples than retaining more solar radio bursts features.
According to the efficiency test, \texttt{ConvRFI} can reach 0.35GB/s processing speed on a laptop with GPU. This makes real time flagging possible for the solar and space weather dynamic spectrum observations of LOFAR.
The method \texttt{ConvRFI} is based on morphological convolution, equivalent to matched filters. For now, considering the features of RFI are usually line-like with sharp edges, we have implemented the corresponding 6 convolutional cores. This can be easily extended to mark other features by changing or appending other convolutional cores accordingly.

Another avenue to the RFI flagging would be to apply machine learning (ML) methods. ML supervised methods are more flexible, because the flagging result of the machine learning models would depend on the training-sets; thus, the model can be fed in a given type of data with given type of RFI feature, and trained to adapt to a particular application scenario. Compared to machine learning based methods, the algorithm-based methods (such as \texttt{AOFlagger, ConvRFI}) have the advantage of less complexity, higher computational efficiency, robustness, and no need for training sets. Thus, algorithmic approaches are widely implemented in the data processing pipelines of radio telescopes like LOFAR and MWA.

\section*{Data Availability}

 
The dynamic spectrum data of LOFAR (ID:L860566,L861370) is publicly available on LTA (\url{http://lta.lofar.org/}) after a period of 1-year data protection according to LOFAR data policy.
The data from recent events are available on request to the author.

\section*{Acknowledgements}

This project is majorly supported by the STELLAR project, which has received funding from the European Union's Horizon 2020 research and innovation programme under grant agreement No 952439. K. Kozarev acknowledges support from the Bulgarian National Science Fund, VIHREN program, under contract KP-06-DV-8/18.12.2019.



\bibliographystyle{mnras}
\bibliography{cite} 




\appendix


\bsp	
\label{lastpage}
\end{document}